\documentstyle[12pt]{article}
\begin{document}
\vspace*{0.5 in}
\begin{center}
\Large
{\bf Atomistic Aspects of Diffusion and Growth  \\
on the Si and Ge (111) Surfaces}
\end{center}
\bigskip \bigskip
\large
\centerline{Efthimios Kaxiras}
\centerline{\it Department of Physics and Divison of Applied Sciences}
\centerline{\it Harvard
University, Cambridge, MA 02138}

\bigskip\bigskip \bigskip \bigskip

\centerline{\bf Abstract}
\medskip

\noindent

The stability of interfaces and the mechanisms of thin film growth
on semiconductors are issues of central importance in electronic devices.
These can only be understood through detailed study of the relevant
microscopic processes.  Experimental studies are able to provide detailed,
atomic scale information for model systems.
Theoretical analysis of experimental results is essential in explaining
certain surprising observations and in providing guidance for
optimizing conditions and methods of growth.
We review recent theoretical work on the diffusion of adatoms,
the structure of adsorbate monolayers, and their implications
for growth on the Si and Ge (111) surfaces.  The
theoretical analysis consists of first-principles calculations
of the total-energy and entropy factors for stable, metastable and
saddle-point configurations.  These calculations are supplemented
by Monte Carlo simulations of simple models that afford direct
contact with experimental observations.

\newpage

\section{Introduction}

The successful production of electronic
devices of increasing complexity and decreasing
size relies on
the ability to control structure down
to exceedingly small scales.
In this context,
understanding the dynamics and the local stability
of {\it atomic} scale structures on semiconductor surfaces becomes
an issue of central importance.
The problem is rendered more complicated by the presence of
adsorbates, which can affect significantly
both the equilibrium geometry and the dynamics of atoms on the surface.
It the last few years impressive advances in
experimental methods have made it possible to study
the motion of atoms on semiconductor surfaces
using probes such as the Scanning Tunneling Microscope (STM)
\cite{Lagally-STM,Williams-STM,Golovchenko-STM,Orr-STM}.
Real-time imaging of the growth process at somewhat
longer length scales has been studied by Low Energy Electron
Microscopy (LEEM) \cite{Tromp-LEEM,Williams-LEEM}.
Perhaps one of the most striking experimental
contributions is the discovery of the
dramatic influence of
adsorbate layers on the nature of growth
of semiconductors \cite{Copel1}.

In this paper we discuss two model systems which have been studied
experimentally in significant detail, and which provide insight
to the issues of adsorbate-modified diffusion and growth on
semiconductor surfaces \cite{adsorbates}.
The first system is the reconstructed Ge(111)-c(2$\times$8) surface
on which a small amount of Pb adatoms promotes atomic diffusion
of a rather unusual type, at room temperature \cite{Golovchenko-STM}.
The second system is the Si(111) surface covered by a full monolayer
of group-V adsorbates, which has been shown to exhibit a
{\it qualitatively}
different  mode of homoepitaxial and heteroepitaxial
growth than the bare substrate \cite{Copel2}.

\section{Diffusion on Ge(111)-c(2$\times$8)}

Several direct measurements of atomic diffusion
on semiconductor surfaces have
been reported using STM \cite{Lagally-STM,Golovchenko-STM}.
One of the most carefully investigated systems is the Si(100) surface,
which has a dimer reconstruction and on which atomic height steps
are ubiquitous.
The long dimer rows and the presence of steps dominate the dynamics
of atoms on this surface, producing highly anisotropic diffusion and
growth patterns in both homoepitaxy and heteroepitaxy.
The situation is qualitatively very similar on the Ge(100) surface.
Much experimental
\cite{Si100steps-Lagally,Si100-steps,Lagally-Ge,SK-GeSi}
and theoretical work
\cite{Si100-Roland,Si100-Metiu,Si100-Sriva,Si100-Rockett,Si100-Brocks,Tersoff}
has been devoted to
understanding diffusion and growth patterns on this system.

The atomic structure, and consequently the dynamics of atoms on (111)
surfaces of Si and Ge are rather different from the (100) surfaces.
Here we will concentrate
on understanding atomic diffusion on the Ge(111) surface,
where recent STM measurements
have revealed some
unexpected findings \cite{Golovchenko-STM}.
The atomic geometry of the (111) surfaces of Si and Ge
is dominated by the adatom
reconstruction.  The basic features of this geometry are shown
in Fig. 1.  Each adatom saturates three surface dangling bonds
by forming covalent bonds to three surface atoms.  In the
equilibrium geometry, the adatom resides directly above a
second layer atom, in the so-called $T_4$ configuration (the
name derives from the position of the adatom being on ``{\it Top}''
of a second layer atom and having 4-fold coordination, if the
second layer atom directly
below is counted as a nearest neighbor \cite{Avouris}).  This structure
induces considerable compressive strain in the immediate
neighborhood of the adatom \cite{Meade}.  The strain is partially compensated
by the presence of surface atoms which are not bonded to
an adatom, the so-called rest atoms.  The simplest geometry
that has low energy and represents a near optimal balance of
electronic energy (saturation of dangling bonds) and strain energy
is a (2$\times$2) reconstruction, in which there is one adatom
and one rest atom in the surface unit cell.  In this reconstruction,
the 3-fold symmetry
of the (111) surface produces a pattern in which every adatom
is surrounded by three rest atoms at equal distances, and vice versa.
This is shown in the upper panel of Fig. 1.
The (2$\times$2) reconstruction in {\it not} observed on
real Si and Ge (111) surfaces, because more
complicated patterns result in a better balance between
electronic and strain energies \cite{Vanderbilt}.
On the Si(111) surface, the observed pattern has a (7$\times$7)
periodicity \cite{Takayanagi},  the unit cell of
which contains local
regions of the (2$\times$2) reconstruction with adatoms and
rest atoms in each half of the unit cell.
The reconstruction
observed on the Ge(111) surface is a slight modification of the
simple (2$\times$2) reconstruction: It consist of a long-range
shift of the relative positions of adatoms in one direction,
giving rise to a c(2$\times$8) pattern.  This does not significantly alter
the local environment, but produces anisotropic diffusion
\cite{Golovchenko-STM}.
For simplicity, we neglect the long-range shift involved
in the c(2$\times$8) reconstruction when we discuss local
single hops of adatoms, and return to it in the full diffusion
simulation, where contact with experiment is made.

In addition to the $T_4$ adatom geometry, there exists
a different geometry which also satisfies the electronic
requirement of reducing the number of dangling bonds on
the surface.  This geometry,
in which an adatom saturates again three surface dangling bonds,
is known as the $H_3$ structure
(the name derives from the fact that the adatom
is above a ``{\it Hollow}'' site, the center of a six-fold ring
formed by first and second layer atoms,
and it is strictly
3-fold coordinated; see Fig. 1, bottom panel).
The $H_3$ structure also introduces compressive strain around the
adatom, which is partially relieved by the surrounding rest atoms.
The balance of electronic and strain energies is less favorable
in this case, making the $H_3$ structure a metastable geometry.
Finally, there exists a simple path for transforming the $T_4$
to the $H_3$ geometry by a single hop of an adatom, which
breaks only one covalent bond.  The saddle-point configuration
for this path is shown in the middle panel of Fig. 1.

In the c(2$\times$8) reconstruction of the Ge(111) surface
all adatoms reside in $T_4$ positions.  The addition of
a small amount (5\% - 10\%)
of Pb adatoms in this system produces interesting effects.
First, the presence of the Pb adatoms allows
diffusion measurements to be made,
because the Pb atoms are larger and their valence
electrons are more weakly bound than the corresponding Ge electrons.
As a result, the Pb atoms appear
as brighter spots than Ge adatoms in the STM pictures and
can be used as the diffusion tracers \cite{Golovchenko-STM}.
Remarkably, the Pb atoms behave very much like Ge atoms,
residing predominantly in $T_4$ adatom positions.  Occasionally they
are also found in $H_3$ positions, whereas this is not the case for
Ge adatoms \cite{Golovchenko-STM}.
The predominant motion seen in experiments involves the
exchange of positions between a regular spot in the STM picture (a
Ge adatom) and a brighter spot (a Pb adatom) at nearest
neighbor sites.
Occasional long jumps are also seen \cite{Golovchenko-STM}.
These results indicate that Pb adatoms
diffuse mostly by exchanging positions with neighboring
Ge adatoms.

In order to establish the viability of the hopping mechanism
shown in Fig. 1 as the elementary diffusion hop, it must
be shown that this mechanism has an activation energy $\epsilon_d$
in agreement with experiment, which gives \cite{Golovchenko-STM} $\epsilon_d =
0.54 \pm 0.03$ eV.  Since the exchange
of positions between adatoms will involve the motion of
both tracers (the Pb adatoms) and regular atoms (the Ge adatoms)
the activation energy for both types of hops must be
calculated.  We have performed such calculations using the first
principles approach outlined in the Appendix.  The results
indicate that the activation energy for the process
shown in Fig. 1, for either Pb or Ge adatoms, is
in agreement with the experimental result:
We find an activation energy of 0.56 eV for Pb adatoms.
For Ge adatoms the activation energy is the same, within
the uncertainty of our calculations (0.01 eV, see Appendix).
This result is within the
the error bar of the experimental activation energy.
It should be emphasized that the present calculations
refer to a layer of adatoms moving
in unison from the $T_4$ to the $H_3$ position in a
(2$\times$2) configuration.
This is different from the actual process of diffusion
on the real surface, where one particular adatom moves
but all the other adatoms remain fixed.
The activation energy associated with the motion
of a single adatom should not be very different from what
we calculated for the motion of a layer of adatoms.
This is because the breaking of covalent bonds is typically
a local disturbance of the system, having little effect
on further neighbors.
Thus, for all practical
purposes, we expect the activation energies quoted above for the
transition from the $T_4$ to the $H_3$ structure of
a full monolayer of adatoms to be good
approximations of the activation energy for hops of
individual adatoms.

Establishing that the mechanism under consideration
has the proper activation energy is only part of the
proof.  It is also desirable to calculate the pre-exponential
factor and make direct comparison with the experimentally
measured diffusion rate.  The diffusion rate is given by
\begin{eqnarray}
D = D_0 e^{-\epsilon_d/k_B T} \\
D_0 = f a^2 \nu e^{S/k_B}
\label{difcon}
\end{eqnarray}
where
$f$ is a geometric factor,
$a$ is the length of the elementary diffusion hop,
$\nu$ is the attempt frequency for elementary hops,
and $S$ is the entropy of diffusion.
For the present case, $a = 2.3$ \AA, the distance between the
$T_4$ and $H_3$ positions on the Ge surface.
If the exchange between nearest neighbors
could be accomplished in a direct way involving only one pair
of adatoms,
then one expects $f = 1$
\cite{fexch}.
The remaining factors need to be calculated explicitly.
For $\epsilon_d$ we use the value quoted above, which
corresponds to the activation energy for single hops
on the real surface and is in excellent agreement with
experiment.  The entropy of diffusion can be calculated
by viewing the hopping process as a reaction between
two states of the system and employing Vineyard's
Transition State Theory (TST) \cite{Vineyard}.  Within this framework
$S$ is given by the following expression:
\begin{equation}
S = k_B \ln \left[ \frac{\int e^{-(E(A)-\epsilon_d)/k_B T} dA}
{\int e^{-E(V)/k_B T} dV} \right]
\end{equation}
This expression applies to a system with $N$ particles
which has a $3N$ configurational space \cite{ceentr}.  The variables
of integration have been rendered dimensionless by
proper scaling.
The integral in the numerator of the logarithm is over the
saddle point surface $A$, a $(3N-1)$ dimensional space separating the
equilibrium and metastable configurations.
The saddle point surface
is defined as being perpendicular to constant energy contours and
passing
through the saddle point configuration.
The integral in the denominator is over
a volume $V$, corresponding to
the full $3N$ dimensional
configuration space on one side of the saddle point surface.
$E(A)$ and $E(V)$ are the total energies of the system
calculated as fucntions of the $3N-1$-dimensional saddle-point surface $A$
and the $3N$-dimensional equilibrium
volume $V$.
Finally a reasonable approximation for
the attempt frequency $\nu$ can be
obtained by the average curvature of the energy surface around the
equilibrium configuration.
In the present case the configuration space for motion of an
adatom on the surface is 2-dimensional.
The remaining atomic coordinates are fixed by the relaxation,
which is taken into account for every position of the adatom.
The saddle-point surface then consists of a 1-dimensional space
that passes through the saddle point.  The
relevant total energy surface,
as obtained from our first-principles calculations, is shown in Fig. 2.
The results are $\nu = 1 \times 10^{11}$ sec$^{-1}$,
and $S = 1.5 k_B$.  This gives
$D_0 = 5 \times 10^{-4}$ cm$^2$ sec$^{-1}$, which
is approximately six orders of magnitude {\it lower} than the
experimental result (see Table I).

In order to understand this very serious discrepancy one
has to take into account the details of the diffusion process
on the real surface.  Specifically, although the $T_4$--to--$H_3$
hop may be the elementary hoping mechanism, it takes
several such hops to achieve a complete exchange between
two adatoms as seen in experiment \cite{Golovchenko-STM}.
Moreover, once an adatom
has performed a single $T_4$--to--$H_3$
hop, it has little choice but to return to its original
position as the next move, unless the neighboring
adatoms also move at the same time and in a coordinated fashion.
This is due to the distribution
of adatom and rest atoms sites, which is illustrated in Fig. 3.
In the equilibrium configuration the adatom is surrounded
by three rest atoms (Fig. 3(a)), and a single
$T_4$--to--$H_3$ hop toward any of
these three positions is allowed.  Once the adatom has
performed a single hop, there are three rest atoms behind it but no
available rest atom in front of it (Fig. 3(b)).  The only allowed
move at this point is in the reverse direction.  This type
of hopping back and forth does not lead to exchange events
with the neighboring adatoms.  The only possibility for
an exchange event to occur is that the neighboring adatoms
happen to move in the right direction, immediately after
the original adatom has moved, thereby opening up
a path for an exchange event.  It is natural to assume
that such events are rather rare, since they depend on the
coordinated motion of several adatoms.
The rarity of such events explains the enormous
factor by which the experimentally measured diffusion is slower
than expected for the rate of single hops.

To describe
the frequency of exchange events quantitatively we have performed
a Monte Carlo (MC) simulation of the hopping process, where
the adatoms are allowed to make uncorrelated $T_4$--to--$H_3$
or $H_3$--to--$T_4$ moves.
The simulation followed the motion of 64 adatoms in
a c(2$\times$8) pattern, consistent with the
actual reconstruction pattern on the Ge surface, with
periodic boundary conditions.
In this simulation the natural
hopping rate is that of the single event.
The forward hop ($T_4$--to--$H_3$) has an activation energy
of $\epsilon_d = 0.56$ eV, as described earlier.  The reverse
hop ($H_3$--to--$T_4$) however, has a lower activation
energy $\epsilon_d'$, because the metastable
configuration has higher energy than the equilibrium
configuration (see Fig. 2).
The relative probability of forward to backward hops
is given by
\begin{eqnarray}
p = e^{-\Delta \epsilon/k_B T} \\
\Delta \epsilon = \epsilon_d - \epsilon_d'
\label{porch}
\end{eqnarray}
Since the forward $T_4$--to--$H_3$
and backward $H_3$--to--$T_4$ hops share a
common saddle point, $\Delta \epsilon$ is also
equal to the energy difference
between the equlibrium and metastable configurations.
If $\Delta \epsilon$
is large compared to $k_B T$, then the adatom will return
very fast to its original position before any
of the neighbors have the chance to move.  If, on the other
hand, this difference is small, the adatom will remain
in the metastable configuration long enough for other
single hopping events to take place on neighboring adatoms.
We find that for Pb adatoms $\Delta \epsilon$ is very small
($\sim 0.02$ eV), whereas for Ge adatoms it approximately an
order of magnitude larger ($\sim 0.2$ eV).
This indicates
that the Pb adatoms will
remain in the metastable
configuration long enough for their neighbors to be displaced
in favorable directions and open up a pathway for the exchange.
In the interest of making the MC simulation faster to
obtain better statistics, one would like the value of $p$
to be as large as is physically plausible.  However, it is not
meaningful to take a very large value of $p$, because then the
adatom pattern quickly becomes disordered.
We find that values of $p$ in the range 0.001 -- 0.01
give reasonable statistics, while the overall ordered c(2$\times$8)
pattern of adatoms
is maintained throughout the simulation,
consistent the experimental results.
For $p$ in that range, the statistical results are not
affected much by its precise value.
A value of $p = 0.005$ corresponds to $\Delta \epsilon = 0.16$
eV at $T = 79$ C (the highest temperature in the experiments).
This value of $\Delta \epsilon$ is consistent
with the difference mentioned earlier between the
forward and backward hop activation energies for Ge adatoms,
which are the majority of moving adatoms (90 - 95 \%).

In the MC simulations we have observed several exchange events.
The events themselves can involve from two to ten adatoms,
which participate in an elaborate ``dance'' around each other,
that allows them eventually to exchange positions.
One such event that involved the motion of five adatoms, aided
by the displacement of three other adjacent adatoms, has
been described in detail elsewhere \cite{Jonah}.
We refer to this mechanism of diffusion as the ``orchestrated
exchange'', to distinguish it from other exchange mechanisms
that have been proposed to describe the motion
of atoms in the bulk \cite{Pandey} or on surfaces \cite{Feibelman}
when strictly two atoms are involved in the exchange event.
Such exchange mechanisms have been called ``concerted exchange''.
Depending on the complexity of the orchestrated exchange event (how many
adatoms are involved), the event can last from
50 (for two adatoms) to 1000 (for ten adatoms) MC moves.
Interestingly, the average separation between events
is $10^{6 \pm 1}$ MC steps!  Fig. 4 displays a portion of our MC
runs with several events shown by spikes, where the
height of the spike represents the number of adatoms
involved in the exchange event.  As is clearly seen from
this figure, the duration of events is negligible compared
to the distance between them.  The results for $p = 0.01$ and
$p = 0.005$ are the same in a statistical sense (same average
duration of events and same average separation between them).
The MC simulation indicates that there is a retardation
in the frequency
of actual exchange events relative to the frequency
of single $T_4$--to--$H_3$ hops equal to $10^{-6 \pm 1}$.
Since this retardation is mainly a geometric effect,
it is convenient to associate it with a very low effective value
for $f$ that appears in Eq.(\ref{difcon}).
This retardation factor brings the results of the calculated
diffusion rate in good agreement with the
experimentally measured diffusion rate (see Table I).

The conclusion to be drawn from the above discussion
is that the mechanism by which adatoms move on a semiconductor
surface at room temperature can be extraordinarily complex.
This results from two factors:
The first factor is that adatoms diffuse
by breaking a minimal number of covalent bonds during each
hop.  In the present case, the elementary hop
breaks only one covalent bond (see Fig. 1).  The
activation energy for breaking a single bond is relatively
low ($\sim 0.5$ eV), making the elementary hop possible at low
temperatures.  Other mechanisms that may correspond to
breaking of several bonds are prohibitively expensive in terms of
activation energy,
and are not seen during low temperature diffusion.
The second factor is that the surface tries to maintain
the adatom reconstruction, which
is its low-energy state.
Diffusion breaks this pattern and is naturally
inhibited by the presence of the ordered structure.
The extreme rarity of orchestrated exchange events
is due to these two factors, since a very complicated sequence of
single hops is required
to achieve exchange, subject to the dual requirements
that each hop breaks only one covalent bond and
that the overall pattern remains a well ordered
reconstruction.

{}From these observations
certain consequences can be inferred about the behavior of the
system under different conditions:
At sufficiently high temperatures,
when the ordered surface pattern disappears, the
diffusion rate should increase substantially since
it will no longer be inhibited by order.  Alternatively,
if the concentration of adatoms could be reduced significantly
at a given temperature,
the diffusion rate should increase by several orders
of magnitude since complicated orchestrated exchanges will no
longer be needed.
It is unclear whether it is possible to
observe these effects experimentally.  Heating up the surface
so as to destroy the ordered pattern may result in adatom
evaporation rather than surface diffusion.  Similarly,
the surface chemical potential, which controls the
concentration of adatoms, may be difficult to change at will
in a real surface.  Real surfaces always contain steps and other
imperfections that act as intrinsic sources or sinks
of adatoms beyond external control.  If a disordered surface
or a lower adatom concentration could be achieved, an increase
in the diffusion rate of adatoms of several (up to six) orders of
magnitude should be observed.

\section{Surfactant mediated growth on Si(111)}

We discuss next the phenomenon of growth on semiconductor
substrates in the presence of monolayers of adsorbates.
It has been shown that the presence of a carefully
chosen adsorbate (called ``surfactant''), can drastically
alter the mode of heteroepitaxial and homoepitaxial growth.
The first demonstration
of this effect on a semiconductor surface was by Copel et al. \cite{Copel1}
for growth of Ge
on the Si(100) surface, using As as a surfactant.
In the absence of the surfactant, only three layers of Ge can
be grown epitaxially on the Si(100) substrate, before 3-dimensional
islands are formed.  In the presence of the surfactant, which
segregates on top of the growing Ge, several dozens of epitaxial Ge layers
can be grown on the substrate.  Similar effects have been
been demonstrated for a variety of other systems including
elementary semiconductors \cite{Copel2,Eaglesham-surf,Horn-Sb},
compound semiconductors \cite{IIIV-surf} and metals \cite{Comsa-Ag,Vegt-Ag},
and using various
surfactants.  More recently, the technique has been used
to improve homoepitaxy of Si on Si, with encouraging results
\cite{Iwanari,Minoda,Nakahara,Voigt,Wilk}.
Theoretical studies of the surfactant effect have
been reported for the Si(100) surface \cite{Oshiyama} and
metal surfaces \cite{Scheffler-Sb,Zhang}.

The use of the word surfactant in the present context is well motivated:
Surfactants are typically agents that reduce the surface
energy.  Since the surfactant must segregate on top of the
growing interface, it is certain that it has lower energy
on the surface rather than embedded below the newly deposited layers.
Thus, a necessary {\it but not sufficient} condition for an adsorbate
to act as a surfactant
is that it lowers the surface
energy of both the substrate and the deposit.
We emphasize that this condition is not sufficient, because the
type of growth depends sensitively
on the kinetics imposed by the presence of the surfactant.  Specifically,
the presence of adsorbate atoms affects the stability of steps, islands,
and other surface features, all of which influence the dynamics
of deposited atoms.

In order to obtain some insight on how these issues play out in
a realistic system, we investigate the case of the Si(111) surface
covered by monolayers of group-V (P, As, Sb) adsorbates.
The group-V elements are natural choices for lowering the
surface energy of Si(111).  The bulk terminated Si(111) surface
consists of a plane of atoms that are three-fold coordinated,
missing a fourth covalent bond to the half crystal that has been
removed.  This is an unstable, high-energy structure for Si,
hence the various reconstructions, e.g., (7$\times$7),
that reduce the density
of dangling bonds and lower the energy.  If the top layer
of Si atoms is replaced by group-V atoms, as shown in Fig. 5(a), a stable
geometry results because the group-V elements prefer
three-fold coordination, forming strong bonds through their
$p$ valence electrons and retaining a pair of electrons in
a filled, low-energy state.
We refer to this geometry as the substitutional geometry.
The substitutional geometry exhibits the periodicity of the
bulk terminated surface, i.e. (1$\times$1).
Alternative group-V atom structures on the
Si(111) surface of potentially low energy can be formed
when one allows for reconstructions of different periodicity.
Two such possibilities
are shown in Fig. 5(b) and 5(c).  The first consists
of group-V atoms bonded in three-atom trimer units, which are
then bonded through covalent bonds to the substrate.
This results in a pattern of periodicity ($\sqrt{3} \times \sqrt{3}$).
The trimer of group-V adatoms can be centered above
a second layer Si atom,
or above the center of a six-fold ring composed of first and second
layer substrate atoms,
as in the case of the single adatom
discussed earlier (see Section II).
The two positions are called the $T_4$
and $H_3$ trimer geometries by analogy to the
single adatom case.  Our first-principles calculations
indicate the the $H_3$ trimer is always higher in energy, so it
will be neglected in the following discussion \cite{Kaxiras-V}.
The second possible structure
consists of one-dimensional chains of group-V atoms
bonded among themselves and to the substrate.
This results in pattern of periodicity (2$\times$1).
In both the trimer and the chain geometry,
every group-V atom has two covalent bonds to other group-V atoms
and one covalent bond to the substrate.  In this respect the
chain and trimer structures are {\it qualitatively} different from the
substitutional structure.
All three geometries, substitutional, trimer and chain,
result in a chemically passive and stable surface layer,
with three-fold coordinated group-V atoms on top.
All Si atoms below this layer have been rendered four-fold coordinated.
The substitutional geometry
has been observed for P and As on Si(111) \cite{Si111-P,Si111-As},
whereas the trimer geometry has been observed
for Sb on Si(111) \cite{Si111-Sb}.

As emphasized above, rendering the substrate chemically stable (thus
lowering the surface energy), is a necessary but not
sufficient condition for surfactant behavior.
It is also desirable that the surfactant affect the kinetics in a
manner favorable to smooth growth.
The first kinetic requirement is that the surfactant can segregate
efficiently
during growth.  By inspecting the three low-energy
geometries for group-V atoms on Si(111) (see Fig. 5),
it is evident that segregating of the surfactants during growth
would be much easier
for the trimer and chain geometries than for the substitutional
geometry.  In the latter geometry, all three covalent bonds
to the substrate have to be severed for the adsorbate
to segregate, whereas in the former two geometries only one covalent bond
per adsorbate atom needs to be severed during growth.

We have performed
extensive first-principles calculations to determine the optimal
geometry for each of the group-V elements \cite{Kaxiras-V}.
The results are summarized in Table II.
In all cases, the trimer and chain geometries are close in energy.
We find that P and As
prefer by a large margin to be in the substitutional geometry,
which is lower
in energy by approximately -0.3 eV per adatom,
compared to the trimer and chain
geometries.
The substitutional geometry
for Sb on the other hand, is energetically unfavorable
compared to the chain or trimer geometries (by approximately
+0.1 eV per adatom).
These results can be rationalized
in terms of the relative strength of chemical bonds between
adsorbate and substrate atoms, and the optimal length of these
bonds which produces strain in the substrate layers \cite{Kaxiras-V}:
The smallest of the three elements, P, introduces compressive
strain in the substitutional geometry,
but produces very strong covalent bonds due to its
chemical affinity to Si.  The middle element, As, fits
comfortably in the substitutional geometry and forms
reasonably strong bonds to Si.
The largest of
the three elements, Sb, cannot be accomodated in the
substitutional geometry and prefers the chain and trimer
structures.  Its chemical affinity to the substrate is
not strong enough to favor the substitutional geometry
in which the number of adsorbate-substrate bonds in maximized.
On a clean Si substrate the Sb-trimer geometry is preferred,
whereas addition of small amount of Ge tilts the balance
towards the chain geometry.
Thus, Sb will form trimers or chains on the substrate.
Both of these Sb structures can easily segregate during growth, whereas
P and As will form the substitutional geometry which is
strongly bound to the substrate and essentially cannot segregate.
Based on these results, we proposed that Sb should be an
effective surfactant on Si(111) but P and As are not
good candidates \cite{Kaxiras-V}.  This prediction has been recently verified
by experimental work \cite{Horn-Sb}.

A full description of the mode of growth in the presence of the
surfactant layer must take into account the actual exchange mechanisms
for the segregation of surfactants on terraces and steps.
This is an exceedingly difficult task, since there
is a multitude of plausible exchange mechanisms, about which
essentially nothing is known at present.  Rather than
identifying all the possibilities and evaluating them
through first-principles calculations -- a daunting if not
an impossible task -- we have proposed a simple solid--on--solid
model that may capture some of the essential features
of surfactant mediated growth \cite{Kaxiras-MSE}.  Here we review the
basic aspects of the model and discuss its implications.

The model is based on the assumption that the substrate
is fully covered by a monolayer of surfactant which always
remains on top of the growing interface, i.e. it segregates
efficiently.  Moreover, diffusion of the new deposits is
very fast on top of the surfactant layer, because the
latter represents a chemically passive environment on
which the deposit cannot bind strongly.
The weak binding leads to small barriers for diffusion
on top of the surfactant.
This indicates that the deposits will travel over long
distances and quickly find any imperfections on
the surface, such as steps.
The diffusion on top
of the surfactant is taken to be free (zero activation energy).
Generally,
the chemical passivation achieved by the surfactant
on terraces is not
as successful in the neighborhood of steps, where
irregular atomic geometries are encountered.
{}From this observation, it is reasonable to assume that
incorporation of the deposits under the surfactant
is facilitated near step edges.  Accordingly, in the
model we considered, atomic exchanges take
place with probability 1 at steps and with probability
$e^{-\epsilon_a/k_B T}$ at terraces, where $\epsilon_a$
is the activation energy for exchange on a flat region
of the surface.
This is an unknown quantity and will be treated
as a parameter in the model.
We have found that, depending on the value of $k_B T/\epsilon_a$,
the surface grows as a smooth film or becomes rough.
To illustrate this we calculate the film roughness, defined by
\begin{equation}
w^2(t) = < [h(\vec{x},t) - \overline{h}(t)]^2 >
\label{wrough}
\end{equation}
where $h(\vec{x},t)$ is the local height at position $\vec{x}$
and $\overline{h}(t)$ is
the average height of the surface, at time $t$.  The angular
brackets in Eq.(\ref{wrough}) denote an average over the
2-dimensional space $\vec{x}$ that spans the substrate.
In Fig. 6 we display $w^2(t)$ for three values of
$k_B T/\epsilon_a$, one below the transition (0.05), one close to
the transition (0.10) and one above the transition (0.20).
$w^2(t)$ is plotted as function of deposition time $t$
which is proportional
to the number of deposited layers (assuming constant deposition rate).
A rough surface corresponds to a steadily increasing $w(t)$, as in
the case of $k_B T/\epsilon_a = 0.20$.
In the other two cases, $w(t)$ fluctuates but remains bounded
for $k_B T/\epsilon_a = 0.05$,
or increases with a very small slope for
$k_B T/\epsilon_a = 0.10$, which is barely visible in the simulation results
shown in Fig. 6.
In the asymptotic regime (growth of a very large number of
layers), values of $k_B T/\epsilon_a$ larger than 0.10 produce
diverging behavior in $w(t)$.  The exact transition point is
somewhat dependent on the size of the simulation cell.
Simulations were performed for square lattices $L \times L$
with $L$ ranging from 100 to 1000, and  periodic boundary conditions.
For these lattices the transition takes place between
$0.1 \leq k_B T/\epsilon_a \leq 0.15$.
A direct picture of the behavior of the model can be given in
terms of the surfaces of films grown in the simulations.
Films corresponding to
$k_B T/\epsilon_a = 0.05, 0.10, 20$,
are shown in Fig. 7(a), (b) and (c)
respectively, for growth of 500 monolayers on a substrate of $L = 100$.
The difference in the film quality in the three cases is striking.
In this particular example, diffusion and exchange at step edges
were made anisotropic by two orders of magnitude in the two
directions to mimic features of the Si(100) surface, the
first system on which surfactant mediated growth was observed \cite{Copel1}.
The anisotropy in diffusion and step-exchange probabilities
produces compact, elongated islands, as seen in experiment.
Applying the model to an isotropic system produces exactly the same
behavior, but the islands have fractal character if step-exchange
at all island edges is equally likely.

The model described here introduces a point of view on the
surfactant effect which is diametrically opposite to
other suggestions. It has been suggested that the
surfactant effect is due to a reduction of the diffusion length
for the new deposits \cite{Tromp-LEEM,Voigt}.
This point of view was advanced in
order to account for the high density of small islands
observed during surfactant mediated growth \cite{Tromp-LEEM,Voigt}.
A high density of small islands naturally leads to
smooth growth by increasing the density of nucleation centers.
What we propose here assumes a very large diffusion
length of new deposits on top of the surfactant,
which allows them to be incorporated predominantly at step edges.
The justification of our point of view is based on
considerations of the atomic structure of the surface
in the presence of the surfactant and its consequences on
the diffusion barriers.
In our model, smooth growth is the result of step flow motion
since all incorporation of new deposits takes place
at island edges.
Nevertheless,
the experimental observation of high density of small
islands needs to be addressed in any physically plausible model.
We note first, that the high density of small islands
is not a universal feature, but is present only for certain
surfactants, such as Sb.  Other surfactants, such as Sn,
result in step flow motion \cite{Iwanari}.

In order to make contact with the experimental results,
and in particular with growth on the Si(111) substrate which is
isotropic, certain modifications in the model
are necessary.  As a first modification we need to ensure
that compact islands are generated even for the isotropic model.
The modification that achieves this
consists of making the step-exchange at island edges
dependent on the coordination:
In a square lattice,
a particle has four nearest neighbors and four next-nearest
neighbors.  If at least two nearest neighbors, or one nearest
neighbor and two next-nearest neighbors, or all four
next-nearest neighbors are island edges, then
the new deposit performs an exchange with the surfactant
atom directly below it
and becomes incorporated in the island with probability 1.
Otherwise the new deposit continues to diffuse on the surface,
or performs an exchange with the surfactant atom
directly below it with probability $e^{-\epsilon_a/k_B T}$,
as on any other flat region of the surface.
We refer to this modified growth model as model A.

To account for the presence of a large density of small islands, we
consider the following situation:  It is possible that
some island edges, and in particular those corresponding to
stable surfactant geometries, are much less likely to lead to
step-exchange between surfactants and new deposits.
The long Sb chains on Si(111) may present
such a configuration due to their intrinsic stability.
Since the long axis of an Sb chain can be oriented
along three equivalent directions on the Si(111) substrate
due to the intrinsic three-fold rotational symmetry,
it is easy to construct islands of triangular shape that
have long chains on each exposed side.  Step-exchange
between deposits and surfactants along these sides
of the islands will be inhibited due to the stability
of the chain.
Other island edges,
which are highly irregular and therefore unstable and chemically active
(such as the ends of Sb chains, for example, seen in Fig. 5(c)),
may be very conducive to step-exchange between
new deposits and surfactants.  This calls for a different
model that can capture these effects.  A simple change in the
features of model A can produce the desired behavior:  If
a new deposit is next to an island edge aligned with one of
the main symmetry directions of the lattice, and if this
edge extends by at least one lattice unit in each direction
from the current position of the deposit, then
the probability of step-exchange at this site is set equal to 0.
All other features of the model are identical to model A.
We refer to this model as model B.

We have performed MC simulations of growth for both models A
and B.  For simplicity, the simulations were carried
out on a square substrate, which should not affect the
qualitative behavior of the models.
Both models lead to smooth growth for small values
of $k_B T/\epsilon_a$.
This is shown in Fig. 8,
where $w^2(t)$ is plotted as a function of deposition time.
The regular oscillations in $w(t)$ correspond to
layer--by--layer growth:  At the lowest value of $w(t)$
an amount of material equal to one monolayer
has been deposited and has been incorporated
essentially in a single layer.  At the highest value of $w(t)$
an additional half a monolayer worth of material has been deposited.
The oscillatory behavior of $w(t)$ continues unchanged for as long as we
run the simulation, up to deposition of several hundreds of monolayers.
For this simulation, the substrate size was chosen as $L = 128$ and
the value of $k_B T/\epsilon_a$ was chosen so that
the probability of exchange at any lattice site away from
island edges is $e^{-\epsilon_a/k_B T} = 2 \times 10^{-4}$.  This choice
of parameters ensures that the diffusion length in the
model, $l_{diff} \sim \sqrt{e^{\epsilon_a/k_B T}}$ is of the same
order of magnitude as the linear dimension of the system.

Interestingly, the type of features
seen upon deposition of a small amount of new atoms on
top of the surfactant is
strikingly different in the two models:
For deposition of 0.1 monolayer,
in model A, one sees
a few very large islands (Fig. 9(a)), whereas in model B a high
density of small islands is found (Fig. 9(b)).
The number of monomers that are not part of larger islands is
approximately the same in the two models, since $k_B T/\epsilon_a$
is the same.
We emphasize that in both models the diffusion length is
essentially infinite (no barrier for diffusion), and the
only difference is the probability of step-exchange
at island edges that have a certain length and orientation.
A more quantitative measure of the difference in the two
models is shown in Fig. 10, where the total area $\Theta (S)$ covered
by islands up to size $S$ is plotted as a function of $S$.
The results of Fig. 10 represent an average taken over 400 samples
for each model, for deposition of 0.1
of a monolayer.
The curves for model A and B are very different for small
island sizes.
The two curves eventually meet at the total deposition of $\Theta = 0.1$.
{}From this figure it is seen that in order to account for 50\%
of the deposited material ($\Theta = 0.05$)
one needs to include islands of size up to $S$ = 70 in model A,
whereas the same amount of deposits is contained in islands
of size only up to $S$ = 7 in model B.
Similarly, in model A islands of
size up to $S$ = 142 are needed to account for 90 \% of the
deposited material, whereas
only sizes up to $S$ = 57 are needed to account for the same
amount of deposited material in model B.
This obviously requires
a much higher number of small islands in model B, as
was illustrated in one particular sample in Fig. 9.
The number of islands of size $S$ per unit area,
$N_S$, is shown in Fig. 11.
In model A there are fewer islands of small sizes and more
islands of large sizes than in model B, as expected from the
discussion above.  The fluctuations in the values of $N_S$
reflect the importance of the island shape in determining
its stability.  Thus, for example, according to the rules
of the simulation, islands of sizes 4, 6, 8, 9, etc.
that can form compact rectangles are very stable in model B,
since all the sides of compact rectangles are places
where step-exchange is inhibited.  In contrast, islands of
sizes 5, 7, etc. which cannot form compact rectangles
are very unstable, so they appear in smaller number densities in model B.
Overall, the density of small islands in model B
compared to model A is overwhelming.

We can interpret the above results in the following manner:
Model A corresponds to a surfactant which does not exhibit
any special behavior as far as the stability of island edges is concerned.
Examples of this kind would be Au or Sn on Si(111), neither of which can
form island edges that are chemically passive and
uncommonly stable.
This is due to the chemical nature of these elements,
which does not allow the formation of
island edges that correspond to saturated covalent structures.
Island edge structures on the Si and Ge (111) surfaces
involve three-fold coordinated sites.
Sn is a tetravalent element which naturally avoids three-fold coordination,
and Au is a metal which cannot induce chemical passivation of
the covalent bonds at island edges.
It has been shown experimentally \cite{Iwanari,Wilk} that
these elements lead to layer--by--layer growth through
step--flow motion, consistent with the presence of few
large islands on the surface.  Model B on the other
hand, corresponds to a surfactant that can form islands
with very stable, chemically passive edges.  One such
example is the long Sb chains discussed earlier (see Fig. 5(c)).
As the simulation shows, in this case one expects
a large density of very small islands to be present
when part of a monolayer has been deposited on top of the surfactant.
Thus, it is not surprising that certain surfactants
produce a large density of small islands, but {\it this does
not necessarily imply a reduction in the diffusion length}.

\section{Discussion and Conclusions}

Through first-principles calculations and diffusion and growth
simulations of the type described above, a detailed
picture of the dynamics of atoms on semiconductor surfaces
is beginning to emerge.
We have shown that at room temperature, adatoms can
move on the Ge(111) surface by performing simple hops
between equilibrium and metastable structures, which
involve the breaking of only one covalent bond.
This corresponds to a relatively low activation energy
of approximately 0.5 eV.  These hops cannot produce long
range transport by themselves, since the density of
adatoms hinders their motion.  Exchanges of adatoms
can only be accomplished through complicated
multi-atom motion.  Such events are very rare in a
surface covered by the equilibrium adatom density.
The frequency of ocurence of such events,
as obtained by our MC simulations of diffusion, is
approximately one million times slower than the single-hop
frequency.
This effective retardation factor provides a natural
explanation of the very slow diffusion rate observed
on the Ge(111)c(2$\times$8) reconstructed surface by STM experiments
\cite{Golovchenko-STM}.
These findings have important implications for
diffusion on other semiconductor surfaces, where
adatoms are present and responsible for mass transport.
The Si(111)(7$\times$7) reconstruction is one relevant example.
We expect that diffusion of adatoms on the
equilibrium reconstructed surface will be rather slow, due to a similar
retardation effect as in the Ge(111)c(2$\times$8) surface.
It is known that upon heating of the Si(111) surface
the (7$\times$7) reconstruction eventually
disappears \cite{Williams-Si111}.
Adatom diffusion on the disordered surface should be
several orders of magnitude faster than in the ordered,
reconstructed surface, while the activation energy of
diffusion in the two cases should be very similar.

In the case of surfactant mediated growth, we find that
elements that are strongly bonded to the substrate (such as
P and As on the Si(111) surface) should be poor surfactant candidates
due to the difficulty they encounter in segregation.  On the other hand,
adsorbates that are weakly bonded to the substrate and form
stable units than can segregate on top of deposits easily
(such as the trimers and chains formed by Sb on the Si(111) surface)
are good surfactant candidates.
These predictions based on our first-principles \cite{Kaxiras-V}
calculations were verified by recent experiments\cite{Horn-Sb}.
Understanding how the presence of surfactants can change
the mode of growth is a more difficult task, requiring
detailed knowledge of kinetic effects.  Since little
is known about atomic exchange mechanisms, we have
instead considered a simple solid--on--solid model and
analyzed its consequences through MC simulations of growth.
The model relies on the assumption that diffusion of the deposits on top
of the surfactant layer is extremely fast and incorporation
takes place mostly at island edges.  The model
exhibits a transition from smooth layer--by--layer growth
to a rough surface when the temperature is raised
beyond a certain value.  Thus, this model suggests that
the presence of surfactants facilitates the
layered growth at low temperature by suppressing nucleation
on top of islands.  Simple modifications of this model
can produce a large density of small islands or a small
density of large islands, depending on specific features
of step-exchange at island edges.  In both cases, the
enhanced diffusion of deposits on top of the surfactant
layer is a key element.  This model represents a point of view
diametrically
different from earlier suggestions that attempted to
explain the surfactant effect in terms of a reduction in
the diffusion length \cite{Tromp-LEEM,Voigt}.
Further study of such models, as well as calculations of
energy barriers of specific exchange mechanism are needed
to provide a full account of the surfactant effect.

The results presented above can also serve as guiding input to treatments of
the growth problem on longer time-scales and length-scales,
e.g. stochastic
growth models involving several rates \cite{Vvedensky},
models based on rate equations \cite{Zangwill},
or statistical mechanical models of surfactant-mediated growth
\cite{Barabasi} that address the asymptotic regime (extremely
large length and time scales).

It is hoped that
the knowledge acquired through the above theoretical studies
will eventually
lead to better methods of semiconductor crystal growth,
in which atomistic structure is exploited to improve
quality and efficiency.

\section*{Acknowledgement}

This work was supported by the Office of Naval Research,
Contract N00014-92-J-1138.  The first-principles
calculations were performed at
the Cornell National Supercomputer
Facility and the simulations at the Aiken Laboratory of the
Division of Applied Sciences, Harvard University.
The help of Jonah Erlebacher in the diffusion simulation
and visualization is gratefully acknowledged.

\section*{Appendix}

The first-principles calculations reported here are based on
density functional theory in the local density approximation
(DFT/LDA) \cite{dft,lda}.
The ions are represented by norm-conserving non-local
pseudopotentials from Bachelet et al. \cite{bachelet},
which make it possible to avoid treating the core electrons
explicitly.
The DFT/LDA formalism provides reliable total-energy comparisons
for a variety of physical systems, including metals, insulators
and semiconductors, as has been shown by extensive applications
over the past two decades \cite{pickett}.
Here, we use the exchange-correlation
functional proposed by Perdew and Zunger \cite{PZexco}, and
a plane wave basis for expanding the electronic wavefunctions.
Convergence of the total energy differences between
different atomic configurations in terms of the number of basis
functions and the sampling of reciprocal space was achieved
by using a basis of plane waves with kinetic energy up to 10 Ry
and special sampling sets \cite{kpts} consisting of up to 16
points in the surface Brillouin Zone.
The surfaces are modeled by slabs which are periodically repeated in the
direction perpendicular to the surface.  The slabs consist of 10
substrate layers separated by vacuum regions equivalent to three
bond lengths of bulk Si.  Inversion symmetry is used to facilitate
the computations and to eliminate charge transfer between the
two sides of the slab.
Full relaxation of the atomic
geometries is included by minimizing the
magnitude of forces on the ions, calculated through the
Hellmann-Feynman theorem.
In the relaxed equilibrium
configurations the forces are smaller than 5 mRy/a.u.
With these computational parameters, we find that
relative energy differences are converged to about 10 meV per
adatom.

\newpage
\begin{center}
{\bf FIGURE CAPTIONS}
\end{center}

\noindent
{\bf Figure 1.}
The adatom geometry on the Si and
Ge (111) surfaces.
The adatom is indicated as a shaded circle.
In the stable position (called $T_4$ geometry, top panel)
the adatom resides
on top of second layer substrate atom.
In the metastable position (called $H_3$ geometry, bottom panel)
the adatom resides
on top of a hollow site, surrounded by a six-fold ring
of first and second layer substrate atoms.
Each adatom is typically surrounded by rest atoms (indicated
by circles with small dots at their center), the relaxation
of which relieves some of the surface strain.
A simple path exists that takes the adatom from the
stable to the metastable geometry.  The saddle point
configuration for this path is shown in the middle panel.

\bigskip
{\bf Figure 2.}
The calculated energy surface for the $T_4$--to--$H_3$
single hop of a Pb adatom on the Ge(111) ($2\times2$) surface.
The energy is given as eV per adatom, and the equilibrium (EQ)
$T_4$, saddle point (SP) and metastable (ME) $H_3$ configurations
are marked.

\bigskip
{\bf Figure 3.}
Illustration of the inhibition of the motion of a single
adatom (marked dark, in the center of the figure) by the presence
of surrounding adatoms (marked by lighter shade).  The three
rest atoms near the central adatom are marked by white circles
with dots.  The remaining white circles represent substrate atoms.
(a) The central adatom in the equilibrium $T_4$ configuration.
(b) The central adatom in the metastable $H_3$ configuration.  Notice the
lack of rest atoms in front of the moving adatom in (b).
If the neighboring adatoms remain fixed, the adatom in (b)
has little choice but to return to its original position shown
in (a).

\bigskip
{\bf Figure 4.}
Results of the MC simulation of diffusion by orchestrated exchange.  The
spikes indicate exchange events.  The height of a spike represents
the number of adatoms involved in an orchestrated exchange event.
The horizontal axis represents time measured in MC steps.
Results for two values of $p$ (see Eq.(\ref{porch})) are shown.
The results for
$p = 0.005$  are shifted vertically by 10 for clarity.

\bigskip
{\bf Figure 5.}
The three lowest energy stable geometries of
group-V atoms on the Si(111) surface.  The Si
substrate atoms are white, the group-V atoms are shaded.
(a) The substitutional geometry of (1$\times$1) periodicity.
(b) The trimer geometry of ($\sqrt{3} \times \sqrt{3}$) periodicity
at the $T_4$ position.
(c) The chain geometry of (2$\times$1) periodicity.

\bigskip
{\bf Figure 6.}
Roughness $w^2(t)$ (see Eq.(\ref{wrough}))
as a function of number of deposited layers ($\sim t$)
for three values of $k_B T / \epsilon_a$.
$w(t)$ rises very fast for
$k_B T / \epsilon_a = 0.20$, rises very slowly for
$k_B T / \epsilon_a = 0.10$, and oscillates but remains bounded for
$k_B T / \epsilon_a = 0.05$.  Results shown here correspond to
a model in which diffusion and step-exchange at islands are
anisotropic by two orders of magnitude in the two main
directions of a square lattice.

\bigskip
{\bf Figure 7.}
(a) Perspective view of 100$\times$100 lattice after deposition of 500 layers
in the anisotropic model,
for $k_B T / \epsilon_a = 0.05$.
(b) Same sa in (a), for $k_B T / \epsilon_a = 0.10$.
(c) Same sa in (a), for $k_B T / \epsilon_a = 0.20$.
The progressively rougher character of the surface, for the
same amount of deposition, is clearly seen.

\bigskip
{\bf Figure 8.}
Same as in Fig. 6, for the isotropic A and B models (see text).
Here $e^{ -\epsilon_a/k_B T} = 2 \times 10^{-4}$.
The oscillatory behavior indicates layer--by--layer growth, which
continues unchanged for several hundreds of monolayers
(to make the oscillations visible, only a small
time interval corresponding to deposition of ten layers is shown).

\bigskip
{\bf Figure 9.}
Typical island distributions for (a) model A, and (b) model B,
for deposition of 0.1 monolayer.  The density of monomers is the
same, while the surface is dominated by few large islands in (a)
and many small islands in (b).

\bigskip
{\bf Figure 10.}
Amount of deposits $\Theta (S)$ included in all islands of
size up to $S$, for models A and B at 0.1 monolayer.

\bigskip
{\bf Figure 11.}
Island density $N_S$ as a function of island size $S$
for models A and B.  The oscillations arise from the
different degree of stability of various islands (see text).
Notice the predominance of small islands in model B.

\newpage

\begin{tabular}{l|c|c|c}  \hline
& Experiment \cite{Golovchenko-STM} & \multicolumn{2}{c}{Theory} \\
&  & DFT/LDA & MC simulation \\
&  & single $T_4$--to--$H_3$ hop & orchestrated exchange \\ \hline
$\epsilon_d$ (eV) & $0.54 \pm 0.03$ & 0.56 & \\
$\nu$ (sec$^{-1}$) &  & 1 $\times 10^{12}$ & \\
$S$ ($k_B$) & & 1.5 & \\
$\lambda$ (\AA) & & 2.3 & \\
$f$ & & 1 & $10^{-6 \pm 1}$ \\
$D_0$ (cm$^2$/sec)
& $1 \times 10^{-9}$
& $5 \times 10^{-4}$ & $5 \times 10^{-10 \pm 1}$ \\ \hline
\end{tabular}

\vspace{4mm}

Table I: Comparison of experimental and theoretical results for
diffusion of adatoms on the Ge(111)c($2\times8$) surface.
The theoretical
results are brought into agreement with experiment
only when the frequency of complicated orchestrated exchange events
is taken into account through MC simulations.

\vspace{10mm}

\begin{tabular}{l|c|c|c|c}  \hline
 Geometry & Periodicity & Si(111):P & Si(111):As & Si(111):Sb  \\ \hline
 Trimer $H_3$ & $\sqrt{3} \times \sqrt{3}$  & 0.16 & 0.13 & 0.09 \\
 Trimer $T_4$ & $\sqrt{3} \times \sqrt{3}$  & 0.0  & 0.0  & 0.0   \\
 Chain & $2 \times 1 $                      & 0.06 &-0.06 & 0.01 \\
 Subsitutional & $1 \times 1 $              &-0.30 &-0.28 & 0.06 \\ \hline
\end{tabular}

\vspace{4mm}

Table II: Relative energies of the different group-V
adsorbate geometries on the Si(111) surface.
The energy of the trimer in the $T_4$ position is
taken as the zero of energy in each case.  The relative
energies are given in eV per adatom.


\begin{thebibliography}{99}
\bibitem{adsorbates} In the present
work we use the word ``adsorbates'' to refer to the foreign atoms
that are specifically introduced in a system in order to alter
the growth mode.  These atoms must be distinguished from the atoms that
will form the desired film on the substrate;  the latter type of atoms
will be referred to
as ``new deposits''.
\end{thebibliography}
\end{document}